\documentclass[preprint,preprintnumbers,amsmath,amssymb]{revtex4}
\usepackage{amsfonts}    
\usepackage{amssymb}
\usepackage{latexsym}
\usepackage{graphicx}
\usepackage[usenames,dvipsnames]{color}

\newcommand{\la}{\lambda}

\newcommand{\De}{\Delta}

\begin{document}

\title{Helium-like and Lithium-like ions: Ground state energy}

\author{Alexander~V.~Turbiner}
\email{turbiner@nucleares.unam.mx}
\author{Juan Carlos Lopez Vieyra}

\email{vieyra@nucleares.unam.mx}
\affiliation{Instituto de Ciencias Nucleares, Universidad Nacional
Aut\'onoma de M\'exico, Apartado Postal 70-543, 04510 M\'exico,
D.F., Mexico{}}

\author{Horacio~Olivares-Pil\'on}
\email{horop@xanum.uam.mx}
\affiliation{Departamento de F\'isica, Universidad Aut\'onoma Metropolitana-Iztapalapa,
Apartado Postal 55-534, 09340 M\'exico, D.F., Mexico}

\begin{abstract}
It is shown that the non-relativistic ground state energy of helium-like and lithium-like ions
with static nuclei can be interpolated in full physics range of nuclear charges $Z$
with accuracy of not less than 6 decimal digits (d.d.) or 7-8 significant digits (s.d.)
using a meromorphic function in appropriate variable with a few free parameters.
It is demonstrated that finite nuclear mass effects do not change 4-5 s.d. for $Z \in [1,50]$ for 2-,3-electron systems and the leading relativistic and QED corrections leave unchanged 3-4 s.d. 
for $Z \in [1,12]$ in the ground state energy for 2-electron system, thus, the interpolation reproduces 
definitely those figures.
A meaning of proposed interpolation is in a construction of unified, {\it two-point} Pade approximant
(for both small and large $Z$ expansions) with fitting some parameters at intermediate $Z$.
\end{abstract}

\maketitle
\section{Introduction}

Let us consider the Coulomb system of the $k$ electrons and infinitely-heavy charge $Z$: $(k\,e, Z)$ with a Hamiltonian
\begin{equation}
\label{H}
  {\cal H}\ =\ -\frac{1}{2} \sum_{i=1}^k \De_i \ -\ \sum_{i=1}^k \frac{Z}{r_i} \ +\
  \sum_{i>j=1}^k \frac{1}{r_{ij}}\ ,
\end{equation}
where $r_i$ is the distance from charge Z to $i$th electron of mass $m=1$ with electron charge $e=-1$,
$r_{ij}$ distance between the $i$th and $j$th electrons, $\hbar=1$. It is universally known that for every $k$
there exists a certain critical charge $Z_c$ above of which, $Z > Z_c$, the system gets bound
forming a $k$ electron ion. We also know that total energy of bound state $E(Z)$ as the function of
$Z$ is very smooth, monotonously-decreasing negative function with the growth of $Z$
eventually approaching the sum of the energies of $k$ Hydrogenic ions.

For two-electron case, $k=2$ (H${}^-$, He, Li${}^+$ etc) with infinitely heavy charge $Z$ (we will call it the {\it static approximation})
the spectra of low-lying states was a subject of intense, sometimes controversial, numerical studies
(usually, each next calculation had found that the previous one exaggerated its accuracy).
This program had run almost since the inception of quantum mechanics \cite{Hylleraas} and continued until 2007 \cite{Nakashima:2007} where the problem was solved for $Z=1 - 10$ for the ground state with overwhelmingly/excessively high accuracy ($\sim 35$ s.d.) from physical point of view.
Recently, it was checked that the energies found in \cite{Nakashima:2007} are compatible with
$1/Z$-expansion up to 12 d.d. for $Z>1$ and 10 d.d. for $Z=1$, see \cite{TL:2016}.
A time ago Nakashima-Nakatsuji made the impressive calculation of the ground state energy of the 3-body problem $(2\,e, Z)$ with finite mass of nuclei \cite{Nakashima:2008}. It was explicitly seen that taking into account the finiteness of the nuclear mass changes in energy the 4th significant digit for $Z=1,2$
and the 5th one for $Z=3 - 10$ (in atomic units). In present paper, inside of the Lagrange mesh
method \cite{Baye:2015} we check and confirm the correctness of all 12 s.d. in both cases of infinite and finite nuclear masses for $Z=1 - 10$ obtained in \cite{Nakashima:2007,Nakashima:2008}; we also calculate ground state energies in both cases of infinite and finite nuclear masses for $Z=11, 12, 20, 30, 40, 50$ with not less than 10 d.d. It is worth mentioning that for $Z=2$ the energy difference for infinite and finite nuclear mass cases, see below Table I, coincides with the sum of the first three orders in mass polarization in 11 d.d. \cite{Pachucki:2017}.

For three-electron case $k=3$ (Li, Be${}^+$ etc) accurate calculations of the ground state energy for $Z=3 - 20$ were carried out in \cite{Drake:1998} for both cases of infinite and
finite nuclear masses. We believe that, at least, 10 s.d. obtained in these calculations are confident. The effect of finiteness of the nuclear mass changes 4th - 3rd decimal digit in the energy (in atomic units) when moving from small to large $Z$.
For $Z = 15 - 20$ (and for infinite nuclear mass) the check of compatibility of obtained results
with $1/Z$-expansion was also made: 5-6 d.d. in energy coincide \cite{Drake:1998}. This coincidence provides
us the confidence to the correctness of the number of decimal digits which is sufficient for our purposes. Note that finite mass effects were found in this case perturbatively,
taking into account one-two terms in the expansion in electron-nuclei reduced mass. We are unaware about any calculations of the ground state energy of the four-body problem $(3\,e, Z)$.

Aim of the present paper is to construct a simple interpolating function for the ground state
energy in full physics range of $Z$ for $k=2,3$ which would provide for ground state energy in the case of infinitely heavy nucleus (the static approximation) not less than 6 d.d. exactly.
Such a number of exact figures is definitely inside of domain of applicability of non-relativistic QED with static nucleus.

As the first step we collect data for the ground state energies available in literature for the cases of both infinite nuclear masses and finite nuclear masses (taking into account the masses of the most stable nuclei, see \cite{Audi:2003}) for two- and three-electron systems, see Table I, II, respectively.
This step is necessary in order to evaluate the effects of finite nuclear mass to the ground state energy:
what significant (decimal) digit in energy is changed.

For $k=2$ the energies for $Z=0.94, 11, 12, 20, 30, 40, 50$ were calculated, see Table I,
employing the Lagrange mesh method \cite{Baye:2015} and using the concrete computer code
designed for three-body studies \cite{Hesse-Baye:2003,OT-PLA:2014}. This method provided systematically the accuracy of 13-14 s.d. for the ground state energy of various 3-body problems~\cite{OT-PLA:2014}.
As for $Z=1 - 10$ the results (rounded to 10 d.d.) obtained in \cite{Nakashima:2007,Nakashima:2008}
are also presented. All these energies were recalculated in the Lagrange mesh method and confirmed in
{\it all} displayed digits in Table I. Note that for non-physical charge $Z=0.94$ we choose the nuclear mass $M_n = 1501.9877 m_e$ following the straightforward interpolation based on of the semi-empirical Bethe-Weizs\"acker mass formula. Taking into account leading relativistic and QED effects obtained in \cite{Yerokhin:2010} for $Z=2-12$ one can see that they leave unchanged the first 3 - 4 s.d. in the ground state energy. Systematically, the finite-mass effects are positive and increase the ground state energy, while the QED and relativistic corrections are negative and tend to decrease the ground state energy.

For $k=3$ (three-electron ions) and infinite nuclear mass the results by Yan et al, \cite{Drake:1998} are presented in Table II. Recently, for $Z=3,4$ they were recalculated by Puchalski et al, \cite{Pachucki:2008} using the alternative method and were confirmed in 9 d.d., while 10-11 d.d. were corrected.
As for finite nuclear mass case for $Z=3-8$ the six d.d. only can be considered as established, except for $Z=8$, see \cite{Drake:1998,Godefroid:2001,Pachucki:2008}. Note that for $Z=3,4$ the sum of the leading QED and relativistic corrections is of the same of order of magnitude as mass polarization but of opposite sign \cite{Pachucki:2008}. They leave unchanged 3-4 s.d. in the ground state energy in static approximation.

{\it Expansions.}\
It is well known that at large $Z$ the energy of $k$-electron ion in static approximation admits the celebrated $1/Z$ expansion,
\begin{equation}
\label{1overZ}
  E(Z)\ =\ -B_0 Z^2 + B_1 Z + B_2 + O\bigg(\frac{1}{Z}\bigg)\ ,
\end{equation}
where $B_0$ is the sum of energies of $k$ Hydrogenic atoms, $B_1$ is the so-called electronic interaction energy, which usually, can be calculated analytically. In atomic units $B_{0,1}$ are rational numbers.
In particular, for the ground state at $k=2$\,\cite{TL:2013},
\[
  B_0^{(2e)}\ =\ 1 \ ,\ B_1^{(2e)}\ =\ \frac{5}{8}\ ,\ B_2^{(2e)}\ =\,  -0.15766642946915\,,
\]
and  $k=3$\,\cite{Drake:1998},
\[
  B_0^{(3e)}\ =\, 9/8               \ ,\ B_1^{(3e)}\ =\, 5965/5832 \ ,
  \ B_2^{(3e)}\ =\, -0.40816616526115\,,
\]
respectively, where $B_2$ is the so-called electronic correlation energy.
The expansion (\ref{1overZ}) for $k=2$ has a finite radius of convergence, see e.g. \cite{Kato:1980}.

In turn, at small $Z$, following the qualitative prediction by Stillinger and Stillinger \cite{Stillinger:1966} and further quantitative studies performed in \cite{TG:2011}, \cite{TLO:2016}, there exists a certain value $Z_B > 0$ for which the energy is given by the Puiseux expansion in integer and half-integer degrees
\begin{equation}
\label{PuiseuxGen}
\begin{split}
 E(Z) =& E_{B} + p_1 \left( Z-Z_{B} \right)
 + q_{{3}} \left( Z- Z_{B}\right)^{3/2} + p_{{2}} \left( Z - {\it Z_B} \right)^{2}
 +q_{{5}} \left( Z- Z_{B} \right) ^{5/2}
\\&
+ p_{{3}} \left( Z- Z_{B} \right)^{3}+q_{{7}} \left( Z- Z_{B} \right)^{7/2}
+ p_{{4}} \left( Z- Z_{B} \right)^{4} + \ldots \ ,
\end{split}
\end{equation}
where $E_{B}=E(Z_{B})$.
This expansion was derived numerically using highly accurate values of ground state energy in close vicinity of $Z > Z_B$ obtained variationally. Three results should be mentioned in this respect for $k=2,3$:
(i) $Z_B$ is {\it not} equal to the critical charge, $Z_B \neq Z_c$,
(ii) the square-root term  $(Z- Z_{B})^{1/2}$ is absent and,
(iii) seemingly the expansion (\ref{PuiseuxGen}) is convergent.
In particular, for the ground state at $k=2$\,\cite{TLO:2016} the coefficients in (\ref{PuiseuxGen}) are,
\[
 Z_B^{(2e)}\ =\ 0.904854\ ,\ E_B^{(2e)}\ =\ -0.407924\ ,\ p_1^{(2e)}\ =\, -1.123470 \ ,
 \]
\begin{equation}
\label{k2par}
 \ q_3^{(2e)}\ =\, -0.197785\ ,\ p_2^{(2e)}\ =\, -0.752842\,,
\end{equation}
while for $k=3$\,\cite{TLO:2016,TLOVN:2017},
\[
 Z_B^{(3e)}\ =\ 2.0090 \ ,\ E_B^{(3e)}\ =\ -2.934281\ ,\ p_1^{(3e)}\ =\ -3.390348\ ,\
 \]
\begin{equation}
\label{k3par}
 q_3^{(3e)}\ =\,  - 0.115425\,, p_2^{(3e)}\ = - 1.101372\,,
\end{equation}
respectively.

{\it Interpolation.}\
Let us introduce a new variable,
\begin{equation}
{\la}^{2}=Z-{Z_B}\,.
\label{lambdadef}
\end{equation}
It can be easily verified that in $\la$ the expansion (\ref{PuiseuxGen}) becomes the Taylor expansion while the expansion (\ref{1overZ}) is the Laurent expansion with the fourth order pole at $\la=\infty$. The simplest interpolation matching these two expansion is given by a meromorphic function
\begin{equation}
\label{Int}
  -\,E_{N,4}(\la(Z))\ =\ \frac{P_{N+4}(\la)}{Q_N(\la)}\ \equiv\ \mbox{gPade}(N+4/N)_{n_0, n_{\infty}} (\la)
  \ ,
\end{equation}
which we call the {\it generalized, two-point Pade approximant}. Here $P, Q$ are polynomials
\[
   P_{N+4}=\sum_0^{N+4} a_k \la^k\ ,\ Q_N=\sum_0^N b_k \la^k\ ,
\]
with normalization $Q(0)=1$, thus, $b_0=1$, the total number of free parameters in (\ref{Int})
is $(2N+5)$. It is clear that $P(0)=E_B$, thus $a_0 = E_B$.
The interpolation is made in two steps: (i) similarly to the Pade approximation theory some coefficients in (\ref{Int}) are found by reproducing exactly a certain number of terms $(n_0)$ in the expansion at small $\la$ and also a number of terms $(n_{\infty})$ at large $\la$-expansion, (ii) remaining undefined coefficients are found by fitting the numerical data, which we consider as reliable, requiring the smallest $\chi^2$. It is a state-of-the-art to choose $(n_0)$ and $(n_{\infty})$.

For both cases $k=2,3$ in (\ref{Int}) we choose $N=4$, which is in a way a minimal number
leading to six decimal digits in fit of energy. It is assumed to reproduce {\it exactly} the first four terms in the Laurent expansion (\ref{1overZ}), $n_{\infty}=4$, and the first three terms in the Puiseux expansion (\ref{PuiseuxGen}), $n_{0}=3$. Thus, we consider the generalized Pade approximant $\mbox{gPade}(8/4) (\la(Z))_{3,4}$.
The remaining six free parameters in
\[
    \mbox{gPade}(8/4) (\la)_{3,4}\ =\ \frac{E_B+a_1 \la+a_2 \la^2+a_3 \la^3+a_4 \la^4+a_5 \la^5+a_6 \la^6+a_7 \la^7+a_8 \la^8}{1+b_1 \la+b_2 \la^2+b_3 \la^3+b_4 \la^4}\ ,
\]
are found making fit. For $k=2$ data from Table I, obtained by Nakashima-Nakatsuji \cite{Nakashima:2007} and via the Lagrange mesh method \cite{OT-PLA:2014},
are fitted. While for $k=3$ data from Table II by Yan et al \cite{Drake:1998} are used.
In Table \ref{Paramsk12} the optimal parameters in $\mbox{gPade}(8/4) (\la)_{3,4}$ for $k=2,3$ are presented.

It is interesting to find from $\mbox{gPade}(8/4) (\la(Z))_{3,4}$ the coefficient in front of $\la^{3}$ in the expansion (\ref{PuiseuxGen}),
\[
  \ q_{3,fit}^{(2e)}\ =\, -0.192510\ ,\ q_{3,fit}^{(3e)}\ =\, -0.09126923\ .
\]
They are quite close to accurate ones in (\ref{k2par}), (\ref{k3par}). In general, expanding the function $\mbox{gPade}(8/4) (\la(Z))$ with optimal parameters, see Table III, around $Z=Z_B$ we get
\begin{equation*}
\begin{split}
E^{(2e)}(Z) \simeq&  -0.4079239753 - 1.123469918 (Z-Z_B)
\\&
- 0.1925102198 (Z- Z_B)^{3/2} - 0.8442237652 (Z-Z_B)^2 + 0.5063843255(Z-Z_B)^{5/2} + \ldots\ ,
\end{split}
\end{equation*}
\begin{equation*}
\begin{split}
E^{(3e)}(Z) \simeq&  - 2.934280640 - 3.390347810(Z-Z_B)
\\&
-0.09126923  (Z-Z_B)^{3/2}  - 1.254645426 (Z-Z_B)^2 + 0.29576206 (Z-Z_B)^{5/2}\ldots\ ,
\end{split}
\end{equation*}
and compare with (4)-(5).

\vskip 1cm

\begin{table}
\caption{\footnotesize
 Helium-like ions\,, the lowest $1s^2\ {}^1S$ state energy:
 for $Z=0.94$ $^{(\star)}$  obtained via the Lagrange mesh method for both infinite and finite nuclear mass, see text;
 for $Z=1 \ldots 10$ \cite{Nakashima:2007} (for infinite nuclear mass,
 it coincides with $1/Z$ expansion, see~\cite{TL:2016}, in all displayed digits) and
 \cite{Nakashima:2008} (finite nuclear mass, it coincides with Lagrange mesh results
 in all displayed digits, see text);
 for $Z=11,12$~\cite{TL:2016} (for infinite mass) and Lagrange mesh results (for
 finite nuclear mass); as 
 for $Z= 20, 30, 40, 50$ the Lagrange mesh results presented for
 both infinite and finite nuclear mass cases;
 for infinite nuclear mass case it is compared with fit (\ref{Int}).\\
 For infinite mass case (2nd column), {\it underlined} digits remain unchanged due to finite-mass effects
 (after its rounding),
 digits given by bold reproduced by fit (7) (after rounding);
 ${()}^{\ast}$ the result of polynomial extrapolation from $Z\in[2-12]$
}
\begin{center}
\begin{tabular}{|c|r|r|r|c|c|}\hline
$Z$ & \multicolumn{4}{|c|}{$E$ (a.u.)}    &Fit  (\ref{Int})  \\
     &  \multicolumn{1}{|c}{Infinite mass} &\multicolumn{1}{c}{Finite mass} & \multicolumn{1}{c}{Difference} & \multicolumn{1}{c|}{Rel.$+$ QED. corr.} &\\
                \hline
0.94$^{(\star)}$
      &  \underline{\bf -0.449}\,{\bf 669}\,043 9    &  -0.449\,353\,763\,3     &  $3.15\times
10^{-4} $  & &   -0.449\,668\,972 \\
 1    &  \underline{\bf -0.527}\,{\bf 751}\,{\bf 01}6 5    &  -0.527\,445\,881\,1     &  $3.05\times
10^{-4} $  & $(-0.06\times 10^{-4})^{\ast}$  &   -0.527\,751\,018 \\
 2    &  \underline{\bf -2.903}\,{\bf 724}\,377 0    &  -2.903\,304\,557\,7     &  $4.20\times
10^{-4}  $ & $-1.12\times 10^{-4}$  &   -2.903\,724\,323 \\
 3    &  \underline{\bf -7.279}\,{\bf 91}3\,412 7    &  -7.279\,321\,519\,8     &  $5.92\times
10^{-4} $  & $-6.76\times 10^{-4}$  &   -7.279\,913\,526 \\
 4    &  \underline{\bf -13.65}{\bf 5}\,{\bf 566}\,{\bf 2}38 4   &  -13.654\,709\,268\,2    &  $0.86\times
10^{-3} $  & $-2.38\times 10^{-3}$  &  -13.655\,566\,09  \\
 5    &  \underline{\bf -22.03}{\bf 0}\,{\bf 971}\,580 2   &  -22.029\,846\,048\,8    &  $1.13\times
10^{-3} $  & $-6.26\times 10^{-3}$  &  -22.030\,971\,42  \\
 6    &  \underline{\bf -32.40}{\bf 6}\,{\bf 246}\,601 9   &  -32.404\,733\,488\,9    &  $0.15\times
10^{-2} $  & $-1.37\times 10^{-2}$ &  -32.406\,246\,55  \\
 7    &  \underline{\bf -44.78}{\bf 1}\,{\bf 445}\,148 8   &  -44.779\,658\,349\,4    &  $0.18\times
10^{-2} $  & $-2.63\times 10^{-2}$  &  -44.781\,445\,14  \\
 8    &  \underline{\bf -59.15}{\bf 6}\,{\bf 595}\,122 8   &  -59.154\,533\,122\,4    &  $0.21\times
10^{-2} $  & $-4.61\times 10^{-2}$  &  -59.156\,595\,13  \\
 9    &  \underline{\bf -75.53}{\bf 1}\,{\bf 712}\,364 0   &  -75.529\,499\,582\,5    &  $0.22\times
10^{-2} $  & $-7.56\times 10^{-2}$  &  -75.531\,712\,32  \\
10    &  \underline{\bf -93.90}{\bf 6}\,{\bf 806}\,515 0   &  -93.904\,195\,745\,9    &  $0.026\times
10^{-1} $  & $-1.17\times 10^{-1}$  &  -93.906\,806\,33  \\
11    &  \underline{\bf -114.28}{\bf 1}\,{\bf 883}\,776 0  & -114.279\,123\,929\,1    &  $0.028\times
10^{-1} $  & $-1.75\times 10^{-1}$  & -114.281\,883\,3   \\
12    &  \underline{\bf -136.65}{\bf 6}\,{\bf 948}\,312 6  & -136.653\,788\,023\,4    &  $0.032\times
10^{-1} $  & $-2.50\times 10^{-1}$  & -136.656\,947\,5   \\[5pt]
20    &  \underline{\bf -387.65}{\bf 7}\,{\bf 23}3\,833 2  & -387.651\,875\,961\,4    &  $5.36\times
10^{-3} $  & & -387.657\,230\,9 \\
30    &  \underline{\bf -881.40}{\bf 7}\,{\bf 3}77\,488 3  & -881.399\,778\,896\,1    &  $7.60\times
10^{-3}  $ & & -881.407\,369\,9 \\
40    &  \underline{\bf -1\,575.15}{\bf 7}\,{\bf 4}49\,525 6 & -1575.147\,804\,148\,0 &  $9.65\times
10^{-3} $  & &  -1\,575.157\,438 \\
50    &  \underline{\bf -2\,468.90}{\bf 7}\,{\bf 4}92\,812 7 & -2468.895\,972\,259\,1 &  $1.15\times
10^{-2} $  & &  -2\,468.907\,478 \\
\hline
\end{tabular}
\end{center}
\label{Interpolation2e}
\end{table}%

\renewcommand{\arraystretch}{0.7} 
\begin{table}
\caption{\footnotesize
 Lithium-like ions\,, lowest, $1s^2\,2s\ {}^2 S$ state energy:
 for $Z=2.16$ $^{(\star)}$ \cite{Yan:2014} (infinite nuclear mass);
 for $Z=3 - 20$ \cite{Drake:1998}  (infinite and finite nuclear mass cases);
 it is compared with fit (\ref{Int}). For $Z=3,4$ finite mass results in second
 row, see ${}^{(\dag)}$, from \cite{Pachucki:2008}.
 For $Z=3\ldots 8$ finite mass results in third-second rows are from \cite{Godefroid:2001}
 with the absolute difference calculated
 with respect to the infinite mass results of \cite{Drake:1998};\\
 for infinite nuclear mass case it is compared with fit (\ref{Int}).\\
 For infinite mass case (2nd column), underlined digits remain unchanged due to finite-mass effects (after its rounding),
 digits given by bold reproduced by fit (7) (after rounding)
}
\begin{center}{\footnotesize
\begin{tabular}{|c|r|r|r|c|}\hline
$Z$ & \multicolumn{3}{|c|}{$E$ (a.u.)}   &Fit  (\ref{Int})  \\
     &  \multicolumn{1}{|c}{Infinite mass} &\multicolumn{1}{c}{Finite mass} & \multicolumn{1}{c|}{Difference} & \\
\hline
  2.16 $^{(\star)}$        &  {\bf -3.478\,108\,3}01\,6   &  &&    -3.478\,108\,26   \\
     3        &\  \underline{\bf -7.47}{\bf 8\,060}\,323\,65    &    -7.477\, 451\, 884\, 70  & $
6.08\times 10^{-4}$  & -7.478\,060\,43   \\
     ${}^{(\dag)}$         &\  \underline{\bf -7.47}{\bf 8\,060}\,323\,91    &    -7.477\, 452\, 121\, 22  & $
6.08\times 10^{-4}$  &                   \\
        &                                      &    -7.477\, 452\, 048\, 02  & $
6.08\times 10^{-4}$  &                   \\
     4        &\qquad \ \underline{\bf -14.32}{\bf 4\,7}{\bf 63}\,176\,47   &   -14.323\, 863\, 441\, 3   & $
9.00\times 10^{-4}$  & -14.324\,762\,7   \\
    ${}^{(\dag)}$          &\qquad \ \underline{\bf -14.32}{\bf 4\,7}{\bf 63}\,176\,78   &   -14.323\, 863\, 713\, 6   & $
8.99\times 10^{-4}$  &                   \\
        &                                      &   -14.323\, 863\, 687\, 1   & $
8.99\times 10^{-4}$  &                   \\
     5        &  \underline{\bf -23.42}{\bf 4\,6}{\bf 0}5\,721\,0    &   -23.423\, 408\, 020\, 3   & $
1.20\times 10^{-3}$  & -23.424\,606\,1   \\
        &                                      &   -23.423\, 408\, 350\, 5   & $
1.20\times 10^{-3}$  &                   \\
     6        &  \underline{\bf -34.77}{\bf 5\,5}{\bf 11}\,275\,6    &   -34.773\, 886\, 337\, 7   & $
1.62\times 10^{-3}$  & -34.775\,511\,4   \\
        &                                      &   -34.773\, 886\, 826\, 3   & $
1.62\times 10^{-3}$  &                   \\
     7        &  \underline{\bf -48.37}{\bf 6\,8}{\bf 98}\,319\,1    &   -48.374\, 966\, 777\, 1   & $
1.93\times 10^{-3}$  & -48.376\,898\,4   \\
        &                                      &   -48.374\, 967\, 352\, 1   & $
1.93\times 10^{-3}$  &                   \\
     8        &  \underline{\bf -64.22}{\bf 8\,5}{\bf 42}\,082\,7    &   -64.226\, 301\, 948\, 5   & $
2.24\times 10^{-3}$  & -64.228\,542\,0   \\
        &                                      &   -64.226\, 375\, 998\, 3   & $
2.17\times 10^{-3}$  &                   \\
     9        &  \underline{\bf -82.33}{\bf 0\,3}{\bf 38}\,097\,3    &   -82.327\, 924\, 832\, 7   & $
2.41\times 10^{-3}$  & -82.330\,337\,9   \\
    10        &  \underline{\bf -102.68}{\bf 2}\,{\bf 23}1\,482\,4   &  -102.679\, 375\, 319\,     & $
2.86\times 10^{-3}$  & -102.682\,232\,   \\
    11        &  \underline{\bf -125.28}{\bf 4\,1}{\bf 9}0\,753\,6   &  -125.281\, 163\, 823\,     & $
3.03\times 10^{-3}$  & -125.284\,190\,   \\
    12        &  \underline{\bf -150.13}{\bf 6\,1}{\bf 9}6\,604\,5   &  -150.132\, 723\, 126\,     & $
3.47\times 10^{-3}$  & -150.136\,196\,   \\
    13        &  \underline{\bf -177.23}{\bf 8\,2}{\bf 3}6\,560\,0   &  -177.234\, 594\, 529\,     & $
3.64\times 10^{-3}$  & -177.238\,236\,   \\
    14        &  \underline{\bf -206.59}{\bf 0}\,{\bf 302}\,212\,3   &  -206.586\, 211\, 017\,     & $
4.09\times 10^{-3}$  & -206.590\,302\,   \\
    15        &  \underline{\bf -238.19}{\bf 2\,3}{\bf 8}7\,694\,1   &  -238.188\, 129\, 642\,     & $
4.26\times 10^{-3}$  & -238.192\,389\,   \\
    16        &  \underline{\bf -272.04}{\bf 4\,4}{88}\,790\,1   &  -272.039\, 780\, 017\,     & $
4.71\times 10^{-3}$  & -272.044\,490\,   \\
    17        &  \underline{\bf -308.14}{\bf 6}\,{\bf 60}2\,395\,3   &  -308.141\, 728\, 192\,     & $
4.87\times 10^{-3}$  & -308.146\,603\,   \\
    18        &  \underline{\bf -346.49}{\bf 8}\,{\bf 7}26\,173\,7   &  -346.493\, 932\, 364\,     & $
4.79\times 10^{-3}$  & -346.498\,730\,   \\
    19        &  \underline{\bf -387.10}{\bf 0}\,{\bf 85}8\,334\,6   &  -387.095\, 367\, 736\,     & $
5.49\times 10^{-3}$  & -387.100\,859\,   \\
    20        &  \underline{\bf -429.95}{\bf 2\,9}{\bf 9}7\,482\,8   &  -429.947\, 053\, 487\,     & $
5.94\times 10^{-3}$  & -429.952\,999\,   \\
 \hline
\end{tabular}
}
\end{center}
\label{Interpolation3e}
\end{table}%

\vskip 1cm

\begin{table}
\caption{Parameters in $\mbox{gPade}(8/4)_{3,4}\,(\la(Z))$ for $k=2,3$ rounded to 8 d.d.,
  3 constraints imposed for the small $\lambda$ limit and 4 constraints for the large $\lambda$ limit.
  For $k=2$ fit done for data corresponding to $Z=0.94, 1, \ldots 10$.  For $k=3$
  the fit done for data corresponding to $Z=2.16, 3,  \ldots 20$.}
\begin{center}
\begin{tabular}{|c|r|r|}\hline
param & $k=2$  &  $k=3$   \\
\hline
         $a_{0}$ &   -0.40792398       &       -2.9342807   \\
         $a_{1}$ &   -1.1766272        &       -3.8825360   \\
         $a_{2}$ &   -3.6426874        &       -11.952771   \\
         $a_{3}$ &   -4.9863349        &       -8.4708298   \\
         $a_{4}$ &   -11.336050        &       -15.768516   \\
         $a_{5}$ &    -7.3954535       &       -6.1294099   \\
         $a_{6}$ &  -14.883559         &       -8.6463108    \\
         $a_{7}$ &   -3.8077114        &       -1.4927915   \\
         $a_{8}$ &   -7.3502129        &      -1.7252376    \\
         $b_{0}$ &   1.0000000         &       1.0000000    \\
         $b_{1}$ &   2.8844275         &       1.3231645    \\
         $b_{2}$ &   6.1757030         &       2.9180654    \\
         $b_{3}$ &   3.8077114         &       1.3269258    \\
         $b_{4}$ &   7.3502129         &       1.5335445    \\
          \hline
\end{tabular}
\end{center}
\label{Paramsk12}
\end{table}%

In Table I and II the results of interpolation for $k=2$ and $k=3$ are presented, respectively.
In general, difference in energy occurs systematically in seventh or, sometimes, in eighth decimal
for all range of $Z$ studied even including unphysical values $Z=0.94$ for $k=2$ and $Z=2.16$ for $k=3$. However, at $k=3$ and $Z>14$ the difference occurs (non-systematically) at one-two portions in
sixth decimal. We do not have an explanation of this phenomenon. It might be an indication to an inconsistency of the variational energies and $1/Z$-expansion found in \cite{Drake:1998}. From other side,
not less than 7-8 significant digits in energies are reproduced exactly in the whole range of physically relevant $Z$ presented in Tables I,II.

Note the analysis of relativistic and QED corrections for two-electron system performed for $Z=2-12$ in \cite{Yerokhin:2010,Pachucki:2017} shows that they are small or comparable with respect to the mass polarization effects for $Z=1,2,3$ and become dominant for large $Z > 3$.
Similar analysis of relativistic and QED corrections of three-electron system, performed for $Z=3,4$ in \cite{Pachucki:2008}, shows that they contribute to the 1st significant digit in the energy difference between infinite and finite mass cases. For both cases of 2- and 3-electron systems the question about the order of relativistic and QED corrections for large $Z$ needs to be investigated. We can only guess that for both systems the domain of applicability of static approximation for any $Z$ is limited by 3 s.d. in the ground state energy.

Concluding we state that a straightforward interpolation between small and large $Z$ in a suitable variable $\la$ (\ref{lambdadef}) based on meromorphic function $\mbox{gPade}(8/4)_{3,4}\,(\la(Z))$ leads to accurate description of 7-8 s.d. of the ground state energy of the Helium-like and Lithium-like ions in static approximation, in $1s^2\ {}^1 S$ and $1s^2\, 2s \ {}^2S$ states, respectively. Interestingly, the simplified interpolation $\mbox{gPade}(5/1)_{2,4}\,(\la(Z))$ with single fitted parameter can reproduced 3-4 s.d. 
in ground state energy for any of both systems in physics range of $Z$, these digits remain unchanged by finite-mass effects. Hence, this interpolation reproduces the ground state energy in static approximation in its domain of applicability.

It seems natural to assume that the similar interpolations have to provide reasonable accuracies for excited states of above systems and even for other many-electron atomic systems. 
It also gives highly accurate results for on-dimensional anharmonic oscillators.
It will be presented elsewhere \cite{TLOVN:2017}. 

Note that a similar {\it two-point} interpolation works extremely well for simple diatomic molecules H$^+_2$, H$_2$ and ${\rm HeH}$ matching perturbation theory at small internuclear distances and multipole expansion, when for the first two systems the instanton-type, exponentially-small contributions at large distances are included. It provides 4-5-6 figures at potential curves at all internuclear distances but 6 figures (!) for energies of rovibrational states \cite{OT:2017}. These results are in domain of applicability of Bohr-Oppenheimer approximation as well as non-relativistic QED.


\begin{acknowledgments}
This work was supported in part by the PAPIIT grant {\bf IN108815} (Mexico).   Also J.C.L.V. thanks PASPA grant (UNAM, Mexico) and the Centre de Recherches Math\'ematiques, Universit\'e de Montr\'eal for the kind hospitality  while on sabbatical leave during which part of this work was done. A.V.T. thanks M.I.~Eides and
V.A.~Yerokhin for useful discussions.
\end{acknowledgments}


\end{document}